\documentclass[english,a4paper]{article}

\usepackage{dcolumn}
\usepackage{bm}
\usepackage{graphicx}
\usepackage{amsmath}
\usepackage{amsfonts}
\usepackage{amssymb}
\usepackage{amsthm}
\usepackage{amstext}
\usepackage{amsbsy}
\usepackage{amsopn}
\usepackage{amscd}
\usepackage{amsxtra}
\usepackage{hyperref}
\usepackage{empheq}
\usepackage{color}

\usepackage{enumitem} 
\usepackage{comment}

\newtheorem{theorem}{Theorem}

\def\C{\mathbb C}
\def\R{\mathbb R}

\def\phi{\varphi}
\def\epsilon{\varepsilon}

\newcommand{\be}{\begin{equation}}
\newcommand{\ee}{\end{equation}}
\newcommand{\ba}{\begin{array}}
\newcommand{\ea}{\end{array}}
\newcommand{\bea}{\begin{eqnarray}}
\newcommand{\eea}{\end{eqnarray}}

\newcommand{\nn}{\nonumber}

\everymath{\displaystyle}

\begin{document}

\title{Characterization of P-divisibility in two-level open quantum systems}

\author{G. Th\'eret\footnote{Laboratoire Interdisciplinaire Carnot de Bourgogne, CNRS UMR 6303, Universit\'{e} de Bourgogne Europe, BP 47870, F-21078 Dijon, France}, C. Lombard-Latune\footnote{Laboratoire Interdisciplinaire Carnot de Bourgogne, CNRS UMR 6303, Universit\'{e} de Bourgogne Europe, BP 47870, F-21078 Dijon, France}, D. Sugny\footnote{Laboratoire Interdisciplinaire Carnot de Bourgogne, CNRS UMR 6303, Universit\'{e} de Bourgogne Europe, BP 47870, F-21078 Dijon, France, dominique.sugny@u-bourgogne.fr}}

\maketitle

\begin{abstract}
We study different characterizations of P-divisibility in two-level open quantum systems whose dynamics are governed by a time-local master equation with time-dependent relaxation rates. Necessary and sufficient conditions for the P-divisibility of the dynamical map are given in terms of inequalities on such relaxation rates. The equivalence between several P-divisibility characterizations existing in the literature is explicitly proven. The connection to the Breuer-Laine-Piilo measure of non-Markovianity is also established. As an application of such characterizations, we study the open dynamics of a qubit interacting with a bosonic mode. More precisely, we characterize the properties of the local map on the qubit generated by its interaction with the bosonic mode, playing the role of an extremely reduced bath. Interesting observations are made, opening perspectives for a deeper physical understanding of CP-divisibility, P-divisibility, and BLP measure.
\end{abstract}



\section{Introduction}

The non-unitary dynamics of a quantum system interacting with its environment is described by the theory of open quantum systems~\cite{breuerbook,weissbook,alickibook,vega2017}. Since in a realistic physical process, a quantum system cannot be perfectly isolated from its surroundings, this theory has long attracted considerable interest in the physics and chemistry communities, and more recently with the rapid development of quantum technologies~\cite{acin,glaserreview,roadmap,kochreview} and quantum thermodynamics~\cite{Vin2016,Kosloff2019,dann1}. It is therefore of utmost importance to characterize such open dynamics generated by the interaction of the system with its environment.

Open dynamics can be described by dynamical maps $\Phi$ that associate the initial state of the system to its state at later times.
Specific properties such as complete positivity (CP)~\cite{choimatrix} or positivity (P) must be satisfied by the dynamical map in order to ensure the physical validity of the reduced state of the system~\cite{carteret2008,pechukas1994,alicki1995,schirmer2004,vacchini2016}. A key concept in the characterization of an open quantum system is its Markovianity, which is related to the absence of memory effects in the dynamics~\cite{Vega2017,wolfprl,rivasprl,RMPbreuer,reviewrivas,tracedistance,andersson,measureother5,reviewPR,Jagadish2020,Jagadish2023}.
While the Markovian behavior in the case of weak coupling is well understood, its non-Markovian counterpart for strong coupling and structured or finite baths is not at the same level of understanding due to the richness and diversity of phenomena. Different definitions of non-Markovianity have been proposed in the literature~\cite{RMPbreuer,reviewrivas,Vega2017,rivasprl}, most of them being based on the trace distance such as the Breuer-Laine-Piilo measure (BLP)~\cite{tracedistance}, on the CP-divisibility, or even on the P-divisibility of the dynamical map~\cite{andersson,measureother5}. Using the dynamical map $\Phi(t,s)$ from times $s$ to $t$ such that $\Phi(t)=\Phi(t,s)\Phi(s)$ with $\Phi(t)=\Phi(t,0)$, we recall that $\Phi$ is CP-divisible (P-divisible) if $\Phi(t,s)$ is CP (P) for all $t\geq s\geq 0$.
Such properties help to characterize the degree of memory effects present in the open dynamics. However, the connection between their formal definitions and the physical parameters characterizing the dynamics of open systems is non-trivial and surprisingly not explicitly made in the literature.

Our main objective here is therefore to provide explicit characterizations of P-divisibility and BLP criterion in terms of the properties of the open dynamics.
We consider open systems described by time-local master equations with time-dependent relaxation rates and frequency transitions~\cite{RMPbreuer,andersson,relaxrates2,wonderen1999,laine2012,megier2020,localnonlocal}, which are time-dependent generalizations of the Gorrini-Kossakovski-Lindblad-Sudarshan (GKLS) form~\cite{lindblad1,lindblad2}.
  Then, we establish necessary and sufficient conditions  for  the  P-divisibility  of  the  dynamical  map  in  terms  of  inequalities  on  the relaxation rates of the master equation. Note that for CP and P, necessary and sufficient conditions have already been established for two-level quantum systems~\cite{maniscalco2007,hall2008,theret2023}.
  We also prove explicitly the equivalence of different P-divisibility criteria for two-level quantum systems. In particular, we show that the general Kossakovski criterion~\cite{lindblad1,RMPbreuer} can be formulated in terms of a more geometric characterization based on the Bloch sphere representation. As a byproduct, we also establish a condition on the relaxation rates to satisfy the BLP criterion.
  As an application of these general results, we study the properties in terms of CP-divisibility, P-divisibility, and BLP criterion, of a qubit coupled to a finite bath composed of a single bosonic mode.

The paper is organized as follows. We describe in Sec.~\ref{sec2} the model system and the time-dependent relaxation rates that characterize its dynamics. Section~\ref{sec3} is dedicated to P-divisibility. We establish necessary and sufficient conditions on the relaxation rates to obtain a P-divisible map. Different criteria of P-divisibility are proposed for two-level quantum systems and we show their equivalence. An application to the study of finite-bath open dynamics is presented in Sec.~\ref{sec5}. Conclusions and prospective views are given in Sec.~\ref{sec6}. The proofs of different theorems are provided in the Appendices. 

\section{The model system}\label{sec2}

We consider an open two-level quantum system interacting with a bath. The state of the system at time $t$ is described  by a density operator $\rho(t)$. The two-dimensional Hilbert space of the system $\mathcal{H}=\mathbb{C}^2$ is spanned by the canonical basis $\{|1\rangle,|2\rangle\}$ associated with the ground and excited states, respectively. We introduce $\mathcal{S}(\mathcal{H})$ the set of density operators, i.e. the set of positive semi-definite Hermitian operators of unit trace and $\mathcal{B}(\mathcal{H})$ the set of Hermitian operators (not necessarily of trace one). The dynamical map $\Phi$ describing the dynamics of the system maps the initial state $\rho(0)$ to $\rho(t)=\Phi(t)\rho(0)$. This map is assumed to be a completely positive trace preserving map (CPTP) to ensure that the properties of the density operator are preserved throughout its time evolution. In the following, we consider a map described by the following GKLS-like master equation
\begin{eqnarray}\label{eq:generalME}
    \frac{d\rho(t)}{dt} &=& -i\left[H(t),\rho(t)\right] \nonumber \\
    & & + \sum_{j=1}^3 \gamma_j(t)\!\left(L_j\rho(t) L_j^\dagger - \frac{1}{2}\left\{ L_j^\dagger L_j, \rho(t)\right\}\right),
\end{eqnarray}
in units where $\hbar=1$ with
$L_1 = \sigma_+ =
\begin{pmatrix}
0 & 1 \\
0 & 0
\end{pmatrix}$,
$L_2 = \sigma_- =
\begin{pmatrix}
0 & 0 \\
1 & 0
\end{pmatrix}$,
$L_3 = \frac{1}{\sqrt{2}}\sigma_z =
\frac{1}{\sqrt{2}}
\begin{pmatrix}
1 & 0 \\
0 & -1
\end{pmatrix}$,
$H(t) = -\frac{\omega(t)}{2}\sigma_z =
\frac{1}{2}
\begin{pmatrix}
-\omega(t) & 0 \\
0 & \omega(t)
\end{pmatrix}$,
and $\sigma_\pm \equiv \frac{1}{2}(\sigma_x \pm i\sigma_y)$, with $\sigma_{x,y,z}$ the Pauli matrices. The diagonal relaxation rates $\gamma_j$ ($j=1,2,3$) can be expressed in terms of the population relaxation rates $\gamma_{12},\gamma_{21}$ and decoherence rate $\Gamma$ as $\gamma_1 = \gamma_{12}$, $\gamma_2 = \gamma_{21}$ and $\gamma_3 = \Gamma - \frac{1}{2}(\gamma_{12} + \gamma_{21})$~\cite{schirmer2004}. We introduce the parameters $\gamma_\pm$ defined by $\gamma_\pm \equiv \gamma_{12}\pm\gamma_{21}$. The necessary and sufficient conditions of CP-divisibility at time $t$, $\gamma_j(t)\geq 0$ ($j=1,2,3$)~\cite{RMPbreuer,reviewrivas,Vega2017}, are then given by
\bea
&&|\gamma_-|\leq \gamma_+,\nn\\
\textrm{and} && 2\Gamma\geq \gamma_+.\label{CP-divcond}
\eea
The density matrix of the system can be written in terms of the Bloch coordinates $(x,y,z)=(\textrm{Tr}[\rho\sigma_x],\textrm{Tr}[\rho\sigma_y],\textrm{Tr}[\rho\sigma_z])$ as
\begin{equation}
\label{corresp}
 \rho
= \frac{1}{2}(\mathbb{I} + x\sigma_x + y\sigma_y + z\sigma_z)
= \frac{1}{2}
\begin{pmatrix}
1+z & x - iy \\
x + iy & 1-z
\end{pmatrix},
\end{equation}
where $\mathbb{I}$ is the identity matrix. We recall that the Bloch ball $\frak{B}$ is the ball in $\mathbb{R}^3$ such that $x^2+y^2+z^2\leq 1$ and a one-to-one mapping to $\mathcal{S}(\mathcal{H})$ can be defined from~\eqref{corresp}. It is then straightforward to show that the GKLS-like equation can be transformed into the Bloch equations
$$
\frac{d}{dt}
\begin{pmatrix}
x \\
y \\
z
\end{pmatrix}
=
\begin{pmatrix}
-\Gamma & \omega & 0\\
-\omega & -\Gamma & 0\\
0 & 0 & -\gamma_+
\end{pmatrix}
\begin{pmatrix}
x \\
y \\
z
\end{pmatrix}
+
\begin{pmatrix}
0 \\
0 \\
\gamma_-
\end{pmatrix},
$$
whose solutions are given by
$$
\left\{
\begin{array}{ll}
x(t) &= e^{-\tilde\Gamma}(x_0\cos\tilde\omega + y_0\sin\tilde\omega)\\
y(t) &= e^{-\tilde\Gamma}(-x_0\sin\tilde\omega + y_0\cos\tilde\omega)\\
z(t) &= s + z_0e^{-\tilde\gamma_+}
\end{array}
\right.,
$$
where $(x_0,y_0,z_0) \equiv (x(0),y(0),z(0))$, $\tilde\Gamma(t) \equiv \int_0^t \Gamma(t')dt'$, $\tilde\gamma_\pm(t) \equiv \int_0^t \gamma_\pm(t')dt'$, $\tilde\omega(t) \equiv \int_0^t \omega(t')dt'$, and $s$ is the solution of $\frac{d}{dt}s(t) = -\gamma_+s(t) + \gamma_-$ such that $s(0) = 0$. Note that $s(t) = e^{-\tilde\gamma_+(t)}\int_0^t e^{\tilde\gamma_+(u)}\gamma_-(u)du $, which depends on $\tilde\gamma_+$ and $\gamma_-$.
Integrating by parts yields $s(t) = \tilde\gamma_-(t) - e^{-\tilde\gamma_+(t)}\int_0^t \tilde\gamma_-(u)e^{\tilde\gamma_+(u)}\gamma_+(u)du$, which now depends on $\tilde\gamma_-$ and $\gamma_+$.
We deduce (by equating both expressions and varying independently $\gamma_\pm$ without changing $\tilde \gamma_\pm$) that the evolution depends only upon the averaged coefficients $\tilde\omega, \tilde\gamma_\pm, \tilde\Gamma$.\\

The associated dynamical map $t\mapsto\Phi(t)$ is given in matrix form by
\begin{equation}
\label{dynmap}
\Phi(t) = \frac{1}{2}
\begin{pmatrix}
1 + s + e^{-\tilde\gamma_+} & 0 & 0 & 1 + s - e^{-\tilde\gamma_+}\\
0 & 2e^{-(\tilde\Gamma - i\tilde\omega)} & 0 & 0\\
0 & 0 & 2e^{-(\tilde\Gamma + i\tilde\omega)} & 0\\
1 - s - e^{-\tilde\gamma_+} & 0 & 0 & 1 - s + e^{-\tilde\gamma_+}\\
\end{pmatrix},
\end{equation}
for a density matrix $\rho$ written as a column vector, $(\rho_{11},\rho_{12},\rho_{21},\rho_{22})^\intercal$, with $\rho_{ij}$ the matrix elements of $\rho$ in the canonical basis.
It will be useful in the following to extend the dynamical map~(\ref{dynmap}) to $\mathcal B(\mathcal H)$ as a map $t \mapsto \Phi(t)\,:\,\mathcal B(\mathcal H) \to \mathcal B(\mathcal H)$. For $q\in\mathcal{B}(\mathcal{H})$, the expression (\ref{corresp}) can be generalized as $q = \frac{1}{2}({\rm Tr}(q)\mathbb{I} + x\sigma_x + y\sigma_y + z\sigma_z)$ where the Bloch coordinates are defined as for $\rho$. This gives in matrix form
\begin{equation}
\label{dynmap2}
\Phi(t)q = \frac{1}{2}
\begin{pmatrix}
{\rm Tr}(q) + z(t) & x(t) - i y(t)\\
x(t) + iy(t) &
{\rm Tr}(q) - z(t)
\end{pmatrix},
\end{equation}
where
$$
\left\{
\begin{array}{cc}
    x(t) = & 2e^{-(\tilde\Gamma - i\tilde\omega)}q_{12}\\
    y(t) = & 2e^{-(\tilde\Gamma + i\tilde\omega)}q_{21}\\
    z(t) = & sq_+ + e^{-\tilde\gamma_+}q_-
\end{array}
\right.
$$
and $q_\pm = q_{11} \pm q_{22}$, the $q_{ij}$ being the coefficients of $q$ in the canonical basis.\\

The density operator $\rho$
 is said to be an \textsl{instantaneous fixed point} of the dynamics at time $t$ if $\dot \rho(t) = 0$, that is,
$$
\begin{pmatrix}
-\gamma_{21} & 0 & 0 & \gamma_{12}\\
0 & i\omega - \Gamma & 0 & 0\\
0 & 0 & -i\omega - \Gamma & 0\\
\gamma_{21} & 0 & 0 & -\gamma_{12}\\
\end{pmatrix}
\begin{pmatrix}
\rho_{11} \\
\rho_{12} \\
\rho_{21} \\
\rho_{22}
\end{pmatrix} = 0.
$$
Using $\rho_{11} + \rho_{22} = 1$, we have $\rho_{11} = \frac{1}{2}\left(1 + \frac{\gamma_{-}}{\gamma_{+}}\right)$ and $\rho_{22} = \frac{1}{2}\left(1 - \frac{\gamma_{-}}{\gamma_{+}}\right)$. We deduce that there is a unique instantaneous fixed point for the dynamics at time $t$ which is the state
$$\rho_{\rm fp} = \frac{1}{2}
\begin{pmatrix}
1 + z_{\rm fp} & 0 \\
0 & 1 - z_{\rm fp}
\end{pmatrix},$$
where $z_{\rm fp}=\frac{\gamma_{-}}{\gamma_{+}}$. When the ratio $\gamma_-/\gamma_+$ is constant, the time evolution of the coordinate $z(t)$ reads
$$
z(t) = z_{\rm fp} - ( z_{\rm fp} - z_0 )e^{-\tilde\gamma_+(t)}.
$$

\section{P-divisibility}\label{sec3}

\subsection{P-divisibility with trace norm}
A first characterization of P-divisibility in terms of trace norm is given  by the following theorem~\cite{measureother5}.
\begin{theorem}
A dynamical map $t \mapsto \Phi(t)\,:\,\mathcal B(\mathcal H) \to \mathcal B(\mathcal H)$ is P-divisible
at time $t$ if and only if, for all non-zero operators $q\in\mathcal B(\mathcal H)$,
\begin{equation}
\label{tracenormcondition}
\frac{d}{dt}{||\Phi(t)q||_1}\leq0,
\end{equation}
where $||A||_1 = \rm{Tr}[\sqrt{A A^\dagger}]$ denotes the trace-norm.
\end{theorem}

\noindent
The trace norm is easily calculated for two-level quantum systems. We have
\begin{align*}
    ||q||_1^2
    &= (|\varphi_+| + |\varphi_-|)^2\\
    &= \left\{
        \begin{array}{ll}
            {\rm Tr}(q)^2 & \textrm{if}\  {\rm det}(q) \geq 0\\
            {\rm Tr}(q)^2 - 4{\rm det}(q) & \textrm{if}\  {\rm det}(q) \leq 0
        \end{array}
        \right.\\
    &= \left\{
        \begin{array}{ll}
            {\rm Tr}(q)^2 & \textrm{if}\  r^2 \leq {\rm Tr}(q)^2\\
            r^2  & \textrm{if}\  r^2 \geq {\rm Tr}(q)^2
        \end{array}
        \right.,
\end{align*}
where $\varphi_\pm$ denote the (real) eigenvalues of $q$ and $r^2 = x^2 + y^2 + z^2$. Using $\textrm{Tr}(q)=\textrm{Tr}(\Phi(t)q)$, we obtain for all $t\geq0$ and all $q\in\mathcal{B}(\mathcal H)$ that
$$
||\Phi(t)q||_1^2 =
\left\{
\begin{array}{ll}
{\rm Tr}(q)^2 & \textrm{if}\  r(t)^2 \leq {\rm Tr}(q)^2\\
r(t)^2  & \textrm{if}\  r(t)^2 \geq {\rm Tr}(q)^2
\end{array}.
\right.
$$
The derivative of the norm $\frac{d}{dt}||\Phi(t)q||_1$ has the same sign as $\frac{d}{dt}||\Phi(t)q||_1^2$.
Hence, condition (\ref{tracenormcondition}) is equivalent to
\begin{equation}
\label{tracenormcondition2}
    \forall q\in\mathcal B(\mathcal H),\ r(t)^2 \geq {\rm Tr}(q)^2 \Rightarrow \frac{dr}{dt}(t)\leq0.
\end{equation}
This condition can be reformulated in terms of the rates $\Gamma, \gamma_\pm$. We get, $\forall q\in\mathcal B(\mathcal H)$,
\begin{equation}
    \label{tracenormcondition3}
    r(t)^2 \geq {\rm Tr}(q)^2 \Rightarrow
    (\Gamma - \gamma_+)z^2 + \gamma_- {\rm Tr}(q) z - \Gamma r^2 \leq 0.
\end{equation}
A drawback of the characterization of P-divisibility by the trace norm (\ref{tracenormcondition}) is that it is given for $q\in\mathcal{B}(\mathcal{H})$, which may not represent a physical state, i.e. such a $q$ may not belong to $\mathcal{S}(\mathcal{H})$ or lie outside the Bloch ball $\frak{B}$. It turns out that the trace-norm condition (\ref{tracenormcondition}) can be refined by considering states $q$ such that, at time $t$, $\Phi(t)q\in\partial\frak{B}$, where $\partial\frak{B}$ denotes the Bloch sphere of pure states, for which $r(t)=1$. Note that we can interpret this condition as if the dynamics were restarted over at time $t$ from the Bloch sphere.

The following result can then be established.
\begin{theorem}
\label{thm:radius}
A dynamical map $t \mapsto \Phi(t)$ is P-divisible
at time $t$ if and only if for any state $q$ such that
$\Phi(t)q \in \partial\frak{B}$,
we have
\begin{equation}
\label{radiusCondition}
\frac{dr}{dt}(t) \leq 0.
\end{equation}
\end{theorem}
This means that, if we consider the dynamical map at time $t\geq0$ and if we apply it to trace-one states $q$ such that $r(t)$ = 1, the radius should not increase locally, i.e. $r(t+dt)\leq 1$.\\

\begin{proof}
First note that condition (\ref{tracenormcondition}) is homogeneous, so we can assume that ${\rm Tr}(q) = 0$ or ${\rm Tr}(q) = 1$.
Hence condition (\ref{tracenormcondition3}) is equivalent to
\begin{equation}
\label{tracenormcondition4}
    \left\{
    \begin{array}{ccc}
        r(t)\geq 1, {\rm Tr}(q) = 1 &\Rightarrow&  (\Gamma - \gamma_+)z^2 + \gamma_- z - \Gamma r^2\leq 0 \\
        {\rm Tr}(q) = 0 &\Rightarrow&  \Gamma (z^2 - r^2) - \gamma_+ z^2 \leq 0
    \end{array}
    \right.
\end{equation}
Clearly, if condition~(\ref{tracenormcondition4}) (or condition~(\ref{tracenormcondition2})) holds, then (\ref{radiusCondition}) holds as well.

Let us show the converse and assume that condition (\ref{radiusCondition}) holds, that is, by taking Eq.~\eqref{tracenormcondition3} with $r(t)=1$ and ${\rm Tr}(q) = 1$, for all $q$ with $\Phi(t)q \in \partial\frak{B}$,
$$
(\Gamma - \gamma_+)z^2 + \gamma_- z - \Gamma \leq 0.
$$
The coordinates $x,y,z$ can be chosen arbitrarily provided that $r = 1$ and the state is pure.
Taking $z=0$ we get $\Gamma\geq0$.
Taking $z = \pm1$ we get $-\gamma_+ + \gamma_- \leq 0$ and $-\gamma_+ - \gamma_- \leq 0$.
By summing these two relations we obtain $\gamma_+ \geq0$ and $|\gamma_-|\leq\gamma_+$. The inequalities $\Gamma(t)\geq0$ and $\gamma_+(t)\geq0$ imply that the last condition in (\ref{tracenormcondition4}) holds using $z^2\leq r^2$.

Consider now $q \in \mathcal B(\mathcal H)$ with ${\rm Tr}(q) = 1$ and $r^2\geq1$.
Since $\Gamma\geq0$, we have
$$
(\Gamma - \gamma_+)z^2  + \gamma_- z - \Gamma r^2 \leq (\Gamma - \gamma_+)z^2  + \gamma_- z - \Gamma
$$
and we know from condition (\ref{radiusCondition}) that $(\Gamma - \gamma_+)z^2  + \gamma_- z - \Gamma \leq 0$ for all $z^2\leq1$.

It remains to study the operators $q$ such that $z^2\geq1$ and $r^2\geq1$.
Since $\Gamma\geq0$ and $z^2\leq r^2$, we have
$$
(\Gamma - \gamma_+)z^2  + \gamma_- z - \Gamma r^2 \leq -\gamma_+ z^2  + \gamma_- z,
$$
with equality if and only if $x=y=0$.\\
If $\gamma_+ = 0$, so is $\gamma_-$ and the inequality is satisfied.
Suppose that $\gamma_+\neq0$.
The right-hand side is a polynomial which is not positive for all $|z|\geq |z_0| = \left|\frac{\gamma_-}{\gamma_+}\right|$.
We know that $|\gamma_-|\leq\gamma_+$, so $|z_0|\leq1$.
Therefore, for any $|z|\geq1$, $-\gamma_+ z^2  + \gamma_- z\leq0$
and condition (\ref{tracenormcondition4}) is  satisfied.
This concludes the proof.\hfill\qedsymbol
\end{proof}

\subsection{P-divisibility in terms of relaxation rates}

Let us now give a characterization of P-divisibility in terms of the rates $\gamma_\pm, \Gamma$.

\begin{theorem}\label{thm:tracenorm}
The dynamical map is P-divisible at time $t$ if and only if
\begin{eqnarray}
\label{pdv1} & & |\gamma_-(t)| \leq \gamma_+(t) 
\nonumber \\
\label{pdv2} \text{and} & & \gamma_-(t)^2 \leq 4\Gamma(t)(\gamma_+(t) - \Gamma(t))~ \textrm{if}\ 2\Gamma(t) \leq \gamma_+(t). \nonumber
\end{eqnarray}
\end{theorem}
\medskip

For the proof see Appendix~\ref{apptrace}. Some comments can be made about Theorem~\ref{thm:tracenorm}. As expected, the conditions for CP-divisibility, i.e. $|\gamma_-|\leq \gamma_+$ and $\gamma_+\leq 2\Gamma$ are stronger than those for P-divisibility. These inequalities also imply that $\gamma_+\geq 0$ and $\Gamma\geq 0$. Using $\gamma_-/\gamma_+ = z_{\rm fp}$, the above conditions can be written as
\begin{eqnarray}
 & & |\gamma_-(t)| \leq \gamma_+(t) \nn \\
 \text{and}& & 2\Gamma(t) \geq \gamma_+(t)(1 - \sqrt{1 - z_{\rm fp}^2})~ \textrm{if}\ 2\Gamma(t) \leq \gamma_+(t) \nonumber,
\end{eqnarray}
i.e. they are very similar to the CP-divisible case~\eqref{CP-divcond} where $2\Gamma(t)\geq \gamma_+(t)$. We observe that we obtain the same condition as CP-divisibility when $z_{\rm fp}^2=1$ or $\gamma_-=\pm \gamma_+$, which corresponds to the situation where the instantaneous fixed point is located on the Bloch sphere.

\subsection{Kossakowski's Condition for P-Divisibility}
Another condition is due to Kossakowski~\cite{kossakowski72,RMPbreuer} and can be expressed as follows for a two-level system.

\begin{theorem}
\label{thm:breuer}
The dynamical map $t \mapsto \Phi(t)$ of a two-level quantum system is P-divisible
at time $t$ if and only if
\begin{eqnarray}
\label{Breuer1}
& & (\gamma_+(t) + \gamma_-(t))|\langle n|\sigma_+|m\rangle |^2 + (\gamma_+(t) - \gamma_-(t))|\langle n|\sigma_-|m\rangle| ^2 \nonumber \\
& & +(\Gamma(t) - \frac{\gamma_+(t)}{2})|\langle n|\sigma_z |m \rangle | ^2\geq0,
\end{eqnarray}
for any orthonormal basis $(| m \rangle,| n \rangle)$ of the Hilbert space $\mathcal H = \C^2$.
\end{theorem}
An explicit proof of this theorem for two-level quantum systems is given in Appendix~\ref{appbreuer}.

\subsection{P-divisibility from local positivity}
We show here that the conditions for P-divisibility can be derived from a local analysis of the positivity of the dynamical map. We recall the following theorem established in~\cite{theret2023}.
\begin{theorem}
The dynamical map $\Phi$ is P if and only if
\begin{eqnarray}
\label{cn1} & & \tilde{\gamma}_+(t) \geq 0, \tilde{\Gamma}(t)\geq 0,~\forall t\geq0, \nonumber \\
\label{cn2} & \text{and}  & s^2 \leq (1 - e^{-\tilde{\gamma}_+})^2~ \textrm{if}\ \tilde{\gamma}_+ \leq 2\tilde{\Gamma}, \nonumber\\
\label{cn3} & \text{and} & s^2 \leq (1 - e^{-2\tilde{\Gamma}})(1 - e^{-2(\tilde{\gamma}_+ - \tilde{\Gamma})}) ~\textrm{if}\ \tilde{\gamma}_+ \geq 2\tilde{\Gamma}. \nonumber
\end{eqnarray}
\end{theorem}
We apply this theorem for a small time $t$. At first order in $t$, we have $\tilde{\Gamma}\simeq \Gamma t$ and $\tilde{\gamma}_+\simeq \gamma_+ t$, where $\Gamma=\Gamma(0)$ and $\gamma_+=\gamma_+(0)$ to simplify the notations. It can also be shown that $s(t)\simeq \gamma_- t$. The first condition gives $\gamma_-^2 t^2\leq \gamma_+^2 t^2$, which leads to $|\gamma_-|\leq \gamma_+$. The second condition is always verified, while the third condition can be transformed into
$$
\gamma_-^2 t^2\leq 2\Gamma t(2(\gamma_+-\Gamma)t)
$$
i.e.
$$
\gamma_-^2\leq 4\Gamma (\gamma_+-\Gamma)
$$
when $2\Gamma\leq \gamma_+$. As might be expected, we find the criteria for P-divisibility from a local study of P, which is a global property in time of the dynamical map.

\subsection{Summary}

\noindent
In summary, we found the following results.

\begin{theorem}[Necessary and Sufficient Conditions for P-Divisibility]
\label{theo6}
Consider an open two-level quantum system.
The dynamics of this system is P-divisible at time $t\geq0$ if and only if one of the following equivalent conditions is satisfied at time $t$.
\begin{enumerate}
\item (Trace-Norm Condition) For all Hermitian non-zero operators $q\in\mathcal B(\mathcal H)$,
\begin{equation*}
\frac{d}{dt}{||\Phi(t)q||_1}\leq0,
\end{equation*}
where $||A||_1 = \rm{Tr}(\sqrt{AA^\dagger}]$ denotes the trace-norm.
\item (Radius Rate Condition)
\begin{equation*}
r(t)=1 \Rightarrow \frac{dr}{dt}(t) \leq 0
\end{equation*}
\item (Kossakowski Condition) For all orthonormal basis $(|m\rangle,|n\rangle)$ of the Hilbert space $\mathcal H$,
\begin{eqnarray*}
& & (\gamma_+(t) + \gamma_-(t))|\langle n|\sigma_+|m\rangle |^2 + (\gamma_+(t) - \gamma_-(t))|\langle n|\sigma_-|m\rangle |^2\\
& & + (\Gamma(t) - \frac{\gamma_+(t)}{2})|\langle n|\sigma_z|m\rangle |^2\geq 0.
\end{eqnarray*}
\item (Relaxation Rate Condition)
\begin{itemize}
	\item[$-$] $|\gamma_-(t)| \leq \gamma_+(t)$;
	\item[$-$] If $2\Gamma(t) \leq  \gamma_+(t)$, then
	$$
	\gamma_-(t)^2 \leq 4\Gamma(t)(\gamma_+(t) - \Gamma(t)).
	$$
\end{itemize}
Note that these inequalities imply $\Gamma(t) \geq 0$ and $\gamma_+(t)\geq 0$.
\end{enumerate}
\noindent
\end{theorem}

\begin{figure}
    \centering
   \includegraphics[width=0.45\textwidth]{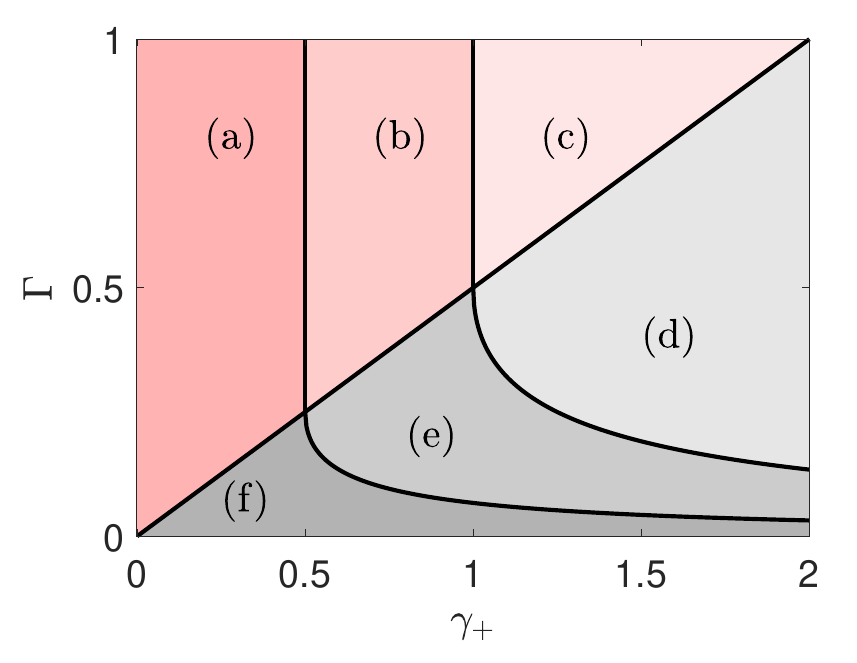}
   \caption{Representation in the space $(\Gamma,\gamma_+)$ of the different domains of CP-divisibility (shades of red) and only P-divisibility (shades of grey). The dynamical map is CP-divisible in the regions (a), (b) and (c) for $\gamma_-=0$, 0.5 and 1, respectively. The map is only P-divisible in the domains (f), (e) and (d) with $\gamma_-=0$, 0.5 and 1, respectively.}\label{fig1}
\end{figure}

The Theorem~\ref{theo6} stating the equivalence between different P-divisibility characterizations is the main result of this paper. A graphical illustration of the properties of the dynamical map with respect to the rates $\Gamma$ and $\gamma_\pm$ is given in Fig.~\ref{fig1}. For a fixed value of $\gamma_-$, we represent in the space $(\gamma_+,\Gamma)$ the different domains for which the dynamical map is CP- divisible or only P- divisible.

Another interesting byproduct of this study is the connection between P-divisibility and the BLP criterion~\cite{tracedistance} which states that there is no back-flow of information from the bath to the system if
$$
\frac{d}{dt}||\Phi(t)(\rho_1-\rho_2)||_1\leq 0
$$
for all initial states $\rho_1$ and $\rho_2$. It is then straightforward to show that the BLP condition is weaker than P-divisibility. According to this criterion, the trace norm of $\Phi(t)q$ should be decreasing only for traceless Hermitian operator $q$. We can deduce a characterization on the relaxation rates of the absence of back-flow of information. Here, condition~\eqref{tracenormcondition3} becomes
$$
\forall q\in\mathcal B(\mathcal H),\ (\Gamma - \gamma_+)z^2 + \gamma_ - \Gamma r^2 \leq 0
$$
or
$$
\forall q\in\mathcal B(\mathcal H),\ -\gamma_+ z^2 - \Gamma (x^2 + y^2) \leq 0,
$$
which gives the sufficient and necessary conditions
$$
\gamma_+ \geq 0,\ \Gamma \geq 0.
$$

\section{Applications}\label{sec5}
\subsection{Standard examples}
We first consider two examples used in the literature to study non-Markovian behaviors.

The first example corresponds to a qubit interacting with multiple decoherence channels~\cite{RMPbreuer,vacchini2012,measureother5}. The Master equation can be written as
$$
\dot{\rho}=\sum_i\frac{\gamma_i}{2}(\sigma_i\rho\sigma_i-\rho)
$$
where $i=x,y,z$ and the $\sigma_i$ are the Pauli matrices. The dynamics can be written in terms of the Bloch coordinates as follows
\begin{eqnarray*}
\dot{x}=-(\gamma_y+\gamma_z)x \\
\dot{y}=-(\gamma_x+\gamma_z)y \\
\dot{z}=-(\gamma_x+\gamma_y)z.
\end{eqnarray*}
We consider the case of eternal non-Markovianity for which $\gamma_x=\gamma_y=1$ and $\gamma_z=-\tanh(t)$~\cite{andersson,megier2017}. We deduce that $\gamma_+=2$, $\gamma_-=0$ and $\Gamma=1-\tanh(t)$. Since $2\Gamma<\gamma_+$ for $t>0$, the map is never CP-divisible. Using $z_{\rm{fp}}=0$, the condition of P-divisibility is $2\Gamma\geq 0$ and the map is always P-divisible. It also verifies the BLP criterion.

The second example is a qubit coupled to a lossy cavity~\cite{RMPbreuer,measurevol}. In this case, the dynamics of the Bloch vector are governed by the following equations
\begin{eqnarray*}
& & \dot{x}=-\frac{\gamma}{2}x+\frac{S}{2}y, \\
& & \dot{y}=-\frac{\gamma}{2}y-\frac{S}{2}y, \\
& & \dot{z}=\gamma-\gamma z.
\end{eqnarray*}
We have $\gamma_-=\gamma_+=\gamma$ and $\Gamma=\frac{\gamma}{2}$. We deduce that $z_{\rm{fp}}=1$. The condition of CP-divisibility and P-divisibility is the same, $2\Gamma\geq \gamma_+$ and is verified at any time $t$.

\subsection{A qubit interacting with a bosonic mode}
In this section, we present an application of the above criteria to a qubit interacting with a single bosonic mode, a rather unusual situation in the open quantum systems community. Traditionally, one uses an explicit microscopic model of a qubit interacting with a bath, and from it one can derive a large variety of master equations depending on the properties of the bath (essentially spectral density and coupling strength with the qubit) and on the approximations used during the derivation of the master equation~\cite{breuerbook}. Here we study a very unusual regime in which the "bath" is reduced to a single bosonic mode. Although a single bosonic mode is definitely not a bath as defined in the standard theory of open quantum system~\cite{breuerbook}, the dynamics induced by the interaction with the bosonic mode is still an open dynamics from the qubit point of view, and moreover can be described by an exact master equation of the same form as~\eqref{eq:generalME}. Thus, even in this unusual configuration, it is legitimate to analyze the properties of the resulting map in terms of CP-divisibility, P-divisibility, and information backflow as defined by the BLP criterion.

The model system can be defined as follows. We consider a two-level system $A$ interacting with a harmonic oscillator $B$ via the Jaynes-Cumming coupling described by the density operator $\rho_{AB}$. The total Hamiltonian can be written as
\be
H_{AB} = \omega_A \sigma_+\sigma_- + g (\sigma_+a + \sigma_- a) + \omega_B a^\dag a.\nonumber
\ee
We assume that at time $t=0$ there is no correlation between $A$ and $B$ and that the initial state of $B$ is in a thermal state at temperature $T_B$, namely $\rho_B(0) = Z^{-1}e^{- \beta_B H_B}$ with $\beta_B=\frac{1}{k_BT_B}$, $H_{B} = \omega_B a^\dag a$ and $Z = {\rm Tr}[e^{-\beta_B H_B}] = (1-e^{-\beta_B\omega_B})^{-1}$. The  reduced dynamics of $A$ given by
\be
\dot \rho_A = {\rm Tr}_B(\dot \rho_{AB}) = {\rm Tr}_B(-i[H_{AB},\rho_{AB}]),\nonumber
\ee
can be written as an exact master equation~\cite{Smirne2010} of the form \eqref{eq:generalME} with
\be\label{Heff}
H(t) = [\omega_A + \omega(t)]\sigma_+\sigma_-,\nonumber
\ee
\be
\omega(t) = - \Im[\frac{\dot\gamma(t)}{\gamma(t)}],\nonumber
\ee
\bea
\gamma(t) &=& \sum_{n=0}^\infty \frac{e^{- n \beta_B\omega_B}}{Z}e^{-i\omega_B t}\nn\\
&&\times\left[\cos(\Omega_n t/2) - i \frac{\Delta}{\Omega_n}\sin(\Omega_n t/2)\right]\nn\\
&&\times\left[\cos(\Omega_{n+1} t/2) - i \frac{\Delta}{\Omega_{n+1}}\sin(\Omega_{n+1} t/2)\right],\nn
\eea
\be
\Omega_n = \sqrt{\Delta^2 + 4 g^2n},\nonumber
\ee
\be\label{gamma1}
\gamma_1(t) = \frac{\alpha(t) \dot\beta(t) - \dot\alpha(t)\beta(t) - \dot\beta(t)}{\alpha(t)+\beta(t) -1},\nonumber
\ee
\be\label{gamma2}
\gamma_2(t) = \frac{\dot\alpha(t)\beta(t) - \alpha(t)\dot\beta(t) - \dot\alpha(t)}{\alpha(t)+\beta(t) -1},\nonumber
\ee
\be\label{eq:alpha}
\alpha(t) = \sum_{n=0}^\infty \frac{e^{- n \beta_B\omega_B}}{Z}\left[ \cos^2(\Omega_n t/2) + \frac{\Delta^2}{\Omega_n^2}\sin^2(\Omega_n t/2) \right],
\ee
\bea\label{eq:beta}
\beta(t) &=& \sum_{n=0}^\infty \frac{e^{- n \beta_B\omega_B}}{Z}\nn\\
&&\times\left[ \cos^2(\Omega_{n+1} t/2) + \frac{\Delta^2}{\Omega_{n+1}^2}\sin^2(\Omega_{n+1} t/2) \right],\nn\\
\eea
\be
\gamma_3(t) =  -\frac{1}{2}\left[\gamma_1(t)+\gamma_2(t) + 2 \Re[\frac{\dot\gamma(t)}{\gamma(t)}]\right],\nonumber
\ee
with $\Delta=\omega_A-\omega_B$. We study below the time evolution of the relaxation rates for three specific situations and we characterize the corresponding properties of the dynamical map.\\

\paragraph{\textbf{A cold mode}.}
The first situation corresponds to the following parameter settings $\omega_B/\omega_A=0.6$, $\Delta/\omega_A = 0.4$, $g/\omega_A=0.3 $, and $\omega_A\beta_B =  2 $. It is said to be \emph{a cold mode} because $\omega_A\beta_B\geq 1$, i.e. the initial temperature of the bath is lower than the typical energy of the two-level system.

\begin{figure}[htpb]
    \centering
   (a) \includegraphics[width=0.4\textwidth]{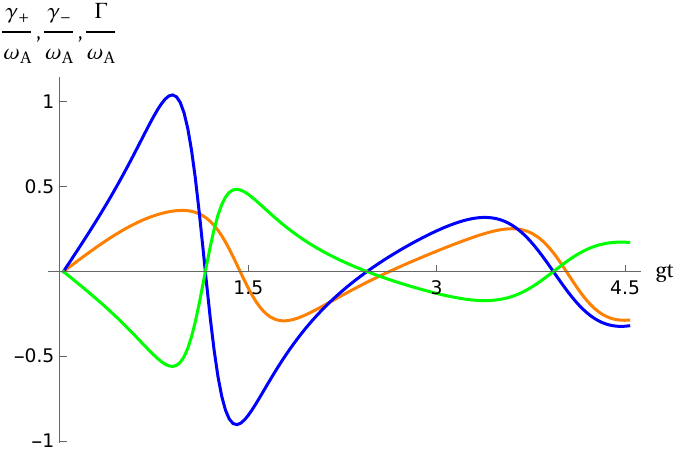}\\
   (b)  \includegraphics[width=0.4\textwidth]{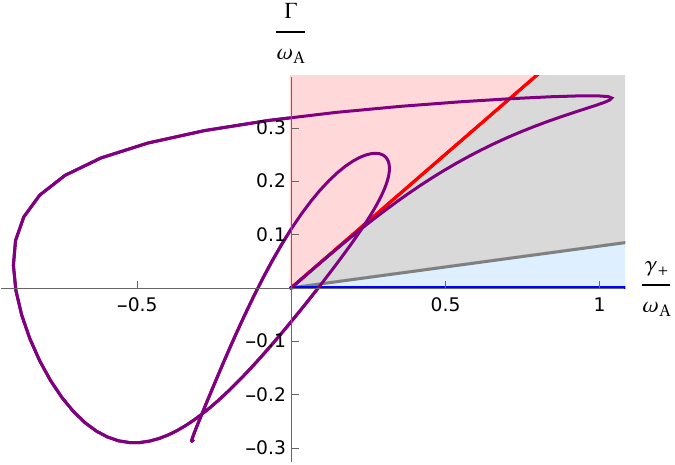}\\\vspace{0.4cm}
   (c)  \includegraphics[width=0.4\textwidth]{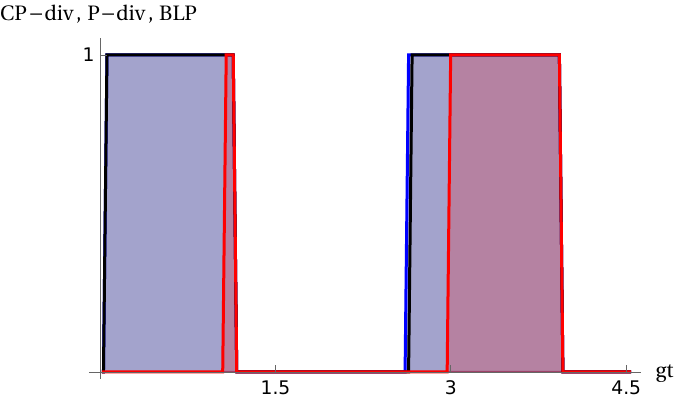}\\\vspace{0.4cm}
   (d) \includegraphics[width=0.3\textwidth]{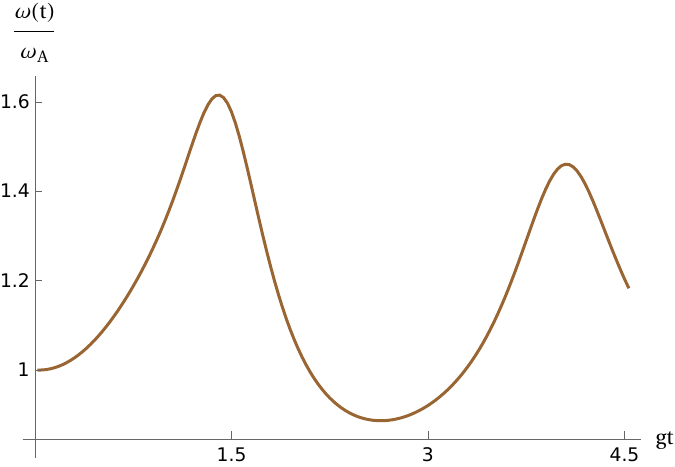}
    \caption{\emph{Cold  mode}: Panel (a) represents the time evolution of $\Gamma(t)$ (in orange), $\gamma_+(t)$, (in blue) and $\gamma_-(t)$ (in green). (b) Evolution of the dynamical map in the space $(\Gamma,\gamma_+)$. The light red, grey and blue areas correspond respectively to the domains of CP-divisibility, P-divisibility only and no information back-flow according to the BLP criterion. The solid red, grey and blue lines are the boundaries between the different regions. The initial point of the trajectory is the origin of the space, $(\Gamma(0)=0=\gamma_+(0))$. (c) Time-dependence of the CP-divisible (red area), P-divisible (grey) and no information back-flow (blue) characters as a function of time. We use a value function that is equal to 1 when the property is true and 0 otherwise. (d) Time evolution of $\omega(t)$.}\label{figcold}
\end{figure}

Figures~\ref{figcold}(a) and (b) represent the time evolution of the relaxation rates. Based on the criteria derived in Sec.~\ref{sec3}, we also show in  Fig.~\ref{figcold}(c)  the time intervals for which the dynamics is CP-divisible, only P-divisible, or with no backflow of information according to the BLP criterion. A visual inspection shows that CP-divisibility leads to P-divisibility, as it should be. We observe that the CP-divisible and P-divisible signatures change very quickly and oscillate with time. Interestingly, this means that although the situation is expected to be highly non-Markovian, since it consists of a qubit strongly interacting with an extremely small bath composed of a single mode, there are still time intervals within which the dynamics is CP-divisible. Another important property common to all regimes is that the ratio  $\gamma_-(t)/\gamma_+(t)$ is constant. This point is shown analytically in Appendix~\ref{app:ratio}. We obtain $\frac{\gamma_-(t)}{\gamma_+(t)} = -\tanh\left(\frac{ \omega_B\beta_B}{2}\right)$. This also implies that the instantaneous fixed point of the dynamics as defined in Sec.~\ref{sec2} does not evolve in time. In the cold mode case, we have $z_{\rm{fp}}=-\tanh(0.6)\simeq -0.537$. Finally, we also plot in Fig.~\ref{figcold}(d) the frequency $\omega(t)$ appearing in the effective Hamiltonian Eq.~\eqref{Heff} and resulting from the strong system-bath interaction. It can be interpreted as work induced by the harmonic oscillator $B$ in the form of unitary energy exchange~\cite{Colla} or non-thermal energy exchange~\cite{Elouard2023}.\\

\begin{figure}
    \centering
   (a) \includegraphics[width=0.4\textwidth]{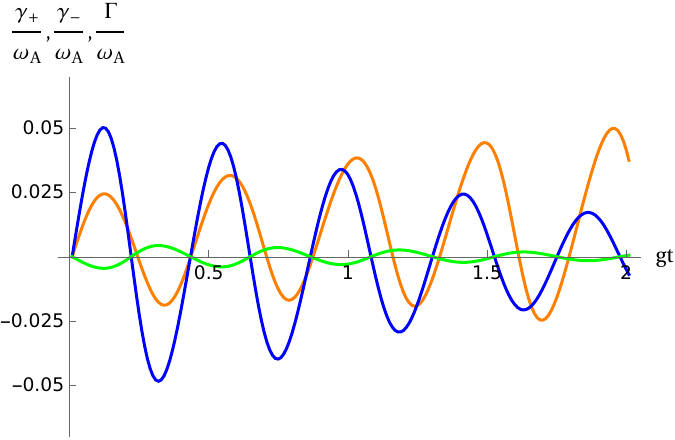}\\
   (b)  \includegraphics[width=0.4\textwidth]{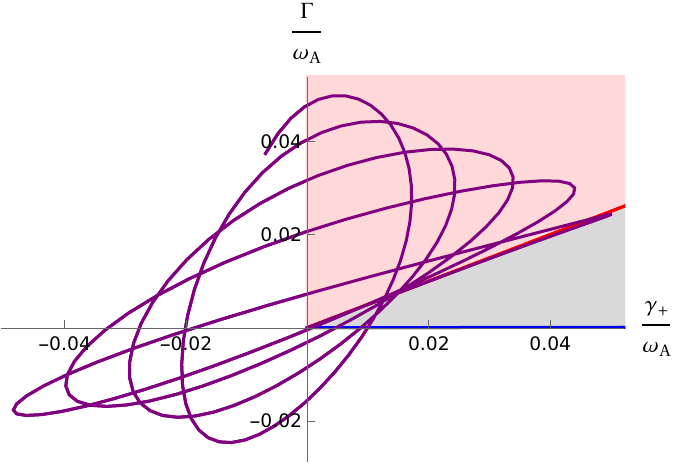}\\\vspace{0.4cm}
   (c)  \includegraphics[width=0.4\textwidth]{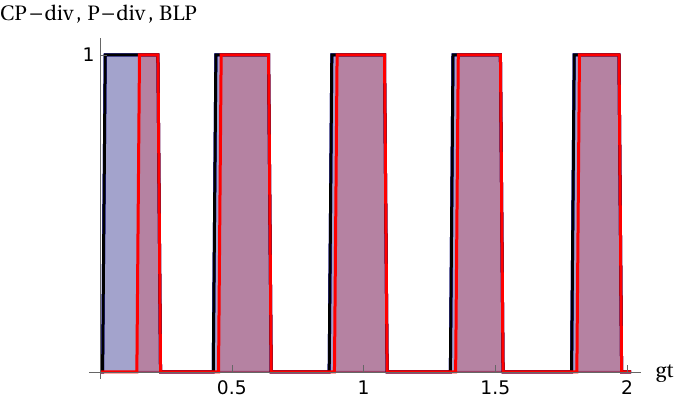}\\\vspace{0.4cm}
   (d) \includegraphics[width=0.3\textwidth]{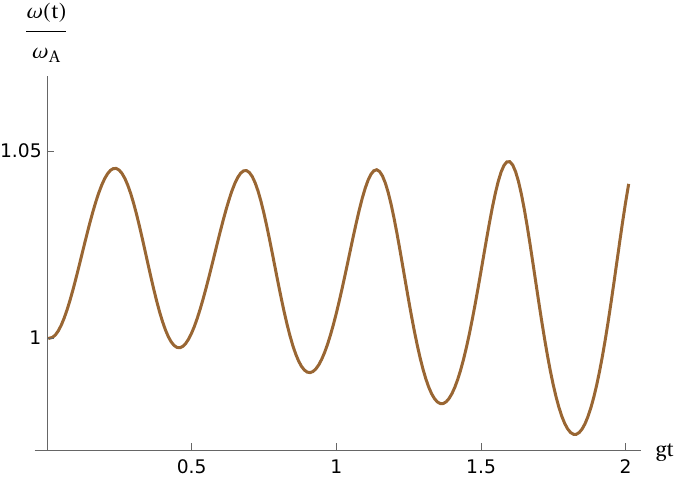}
    \caption{Same as Fig.~\ref{figcold} but for a hot mode.}\label{fighot}
\end{figure}

\paragraph{\textbf{A hot mode}.}
The same analysis is performed in a second situation, referred to as "hot mode" configuration, which corresponds to $\omega_B/\omega_A=0.6$, $\Delta/\omega_A = 0.4$, $g/\omega_A=0.03 $, and $\omega_A\beta_B = 0.3 $. Hot means here that the initial temperature of the bath is larger than the typical energy of the system, i.e. $\omega_A\beta_B<1$. Figures~\ref{fighot}(a) and (b) present the evolution of the relaxation rates in this regime. We observe that their dynamics are much more oscillating than in the first case. As can be seen in Fig.~\ref{fighot}(b) and (c), the dynamical map goes very quickly from the CP-divisible area to the P-divisible domain and regions of information back-flow. In the hot mode case, we have $z_{\rm{fp}}=-\tanh(0.09)\simeq -0.0898$, i.e a fixed point closed to the center of the Bloch ball. We observe that the CP- and P-divisible characters of the dynamical map are almost identical for all times, except initially. Additionally, as in the cold mode situation, we plot in Fig.~\ref{fighot}(d) the frequency $\omega(t)$ appearing in the effective Hamiltonian Eq.~\eqref{Heff}. This time, the oscillations of the bath-induced effective energy are small compared to $\omega_A$, as a consequence of a smaller system-bath coupling.



\paragraph{\textbf{The weak coupling regime}.}

\begin{figure}
    \centering
   (a) \includegraphics[width=0.4\textwidth]{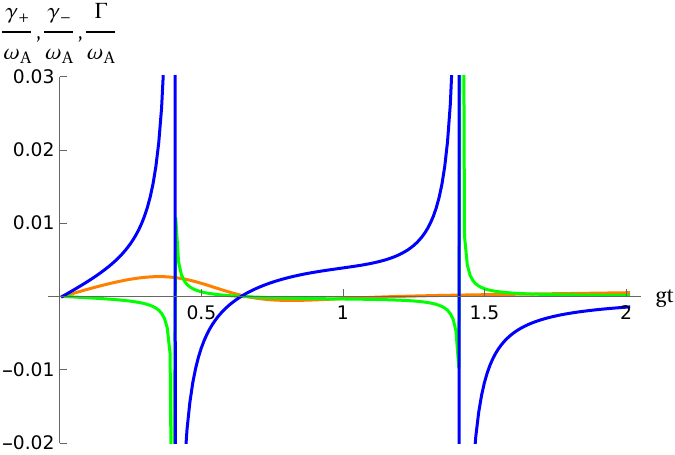}\\
   (b)  \includegraphics[width=0.4\textwidth]{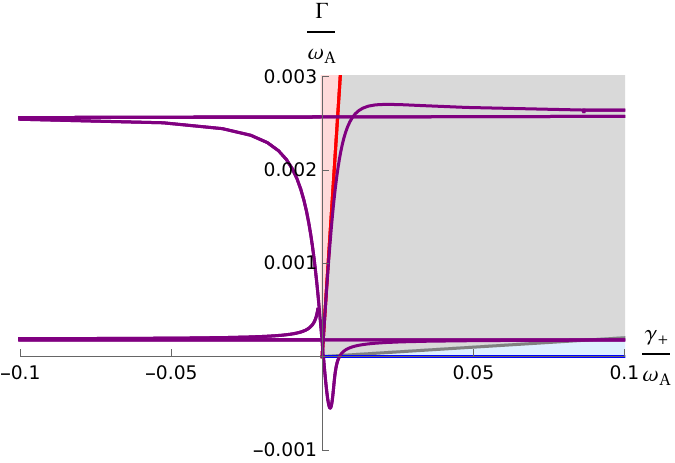}\\\vspace{0.4cm}
   (c)  \includegraphics[width=0.4\textwidth]{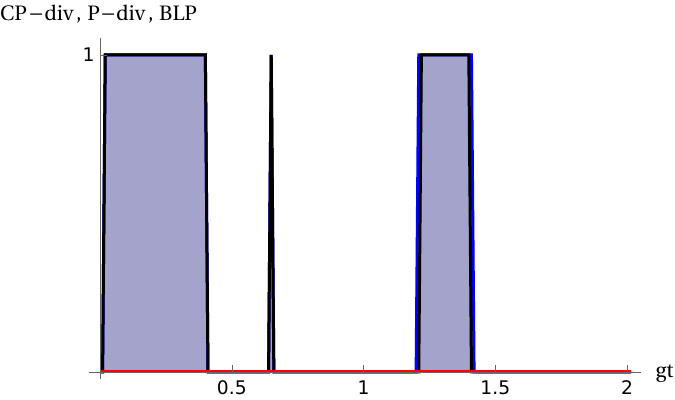}\\\vspace{0.4cm}
   (d) \includegraphics[width=0.3\textwidth]{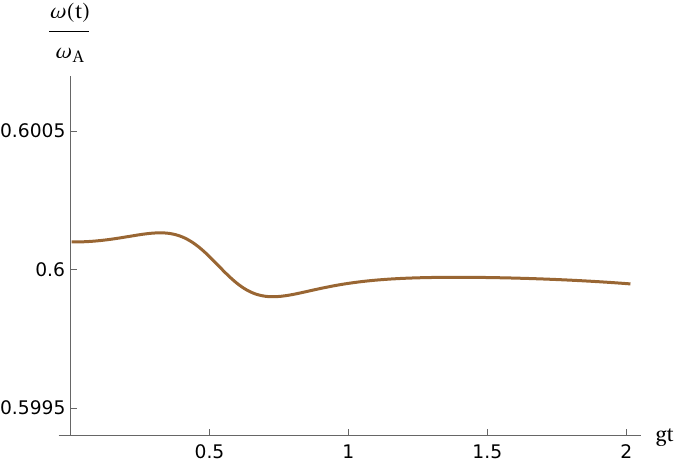}
    \caption{Same as \ref{figcold} but in the weak coupling regime.}
    \label{figweak}
\end{figure}

In the third case referred to as "weak coupling" configuration, the parameters are set to  $\omega_B/\omega_A=0.6$, $\Delta/\omega_A = 10^{-4}$, $g/\omega_A=10^{-3}$, and $\omega_A\beta_B = 0.3 $. This example is characterized by a very low coupling strength. Since the position of the instantaneous fixed point of the dynamics does not depend on the coupling, we have as above for the "hot mode" case $z_{\rm{fp}}=-\tanh(0.09)\simeq -0.0898$. In this weak coupling regime, we observe in Fig.~\ref{figweak}(a) a divergence of $\gamma_+(t)$ (and consequently of $\gamma_-(t)$) around $g t \simeq 0.4$ and $g t \simeq 1.4$.  This is a curious situation with an infinite energy decaying rate ($\gamma_+(t)$), while the decoherence rate ($\Gamma(t)$) remains finite and small.
This feature appears when the qubit and the harmonic oscillator are close to resonance ($\Delta \ll \omega_A, \omega_B$), and curiously still occurs in weak coupling situations.



Additionally, one can see in Fig.~\ref{figweak}(d) that the plot of $\omega(t)$ exhibits an almost constant behavior, which we interpret as a result of weak coupling. It implies that no work is transferred to the qubit, as would be expected in a weak coupling configuration.

Finally, there is an important characteristic common to all situations. One can see in Figs.~\ref{figcold}(b), \ref{fighot}(b) and \ref{figweak}(b) that at short times the trajectory in the $(\Gamma,\gamma_+)$ space is tangent to the frontier between the CP-divisibility and P-divisibility zones, i.e. the line $\Gamma = \gamma_+/2$. Indeed, one can show analytically that $\Gamma(t) - \gamma_+(t)/2 = {\cal O}(t^3)$, for all $g$, $\Delta$, and $\beta_B$. Then, in this limit, one can consider the dynamics to be CP-divisible if higher-order terms are neglected.

\section{Conclusion}\label{sec6}
In this paper, we provide different characterizations of P-divisibility for two-level quantum systems and show their equivalence. Interestingly, one of the signatures is given in terms of inequalities on the relaxation rates and can be easily verified when the dynamics is governed by a time-local master equation. The connection with other dynamical properties such as CP-divisibility and the BLP criterion is also discussed. A hierarchy in the P-divisibility at a given time $t$ is related to the position of the instantaneous fixed point of the dynamics (see Fig.~\ref{fig1}). If the latter is on the Bloch sphere, then P-divisibility is equivalent to CP-divisibility, while for a fixed point at the center of the Bloch sphere, i.e. for unital dynamics, P-divisibility corresponds to the BLP criterion. In all other cases, the different concepts are not equivalent. Such correspondences are not fixed in time and evolve with the relaxation rates of the dynamical map.

This study paves the way to other interesting problems. A natural question is to generalize the relaxation rate criterion established in this paper to more complicated systems with three or more levels~\cite{schirmer2004,coherencevector}. In this case it would also be valuable to characterize the different notions of $n$-divisible maps introduced in~\cite{measureother5}. Another related question is based on the relative entropy between arbitrary pair of states, which is decreasing in time if the map is P-divisible~\cite{AIP}. The violation of this condition indicates that the map is not  P-divisible. In this direction, one can wonder if another concept of relative entropy could be defined with properties that are satisfied for any P-divisible map. A final question concerns the optimal control of open quantum systems~\cite{kochreview,ansel2024} and its connection with the CP- of P-divisibility of the dynamical map. For example, one can ask whether such concepts play a role in the set of states that can be reached by the system and in the structure of the associated control protocol~\cite{lapert2013,reich2015,poggi2017,mangaud2018,atabek2018,mukherjee2015}.

Finally, the application of such criteria to "extremely small bath" situations reveals some unexpected features, and raises the question of a deeper physical understanding of such central concepts as CP-divisibility and P-divisibility.\\

\noindent \textbf{Acknowledgements.}\\
C.LL. acknowledges funding from the French National
Research Agency (ANR) under grant ANR-23-CPJ1-
0030-01.

\appendix

\section{Proof of Theorem~\ref{thm:tracenorm}}\label{apptrace}
\begin{proof}
We prove Theorem ~\ref{thm:tracenorm} by using  condition (\ref{radiusCondition}) given in Theorem \ref{thm:radius}.
The dynamical map is P-divisible if and only if, for all trace-one operators $q$ such that $r(t)=1$, we have $\frac{dr^2}{dt}(t) \leq 0$, that is, at time $t$,
\begin{equation}
\label{radiusCondition2}
z^2(\Gamma - \gamma_+) + z\gamma_- - \Gamma \leq 0.
\end{equation}
Since $r=1$, we have $-1\leq z \leq +1$.
We study the extremum of the polynomial $R(z) = z^2(\Gamma(t) - \gamma_+(t)) + z\gamma_-(t) - \Gamma(t)$ over $[-1,+1]$ (Here $t$ is fixed and we will drop it in the following).\\
Assume that $\Gamma = \gamma_+$.
Taking $|z|=1$, condition (\ref{radiusCondition2}) is satisfied if and only if $|\gamma_-|\leq\gamma_+$ (and in particular we have $\gamma_+\geq0$ and $\Gamma\geq0$).\\
Suppose now $\Gamma \neq \gamma_+$. The extremum of $R$ is attained for $z_m = \frac{-\gamma_-}{2(\Gamma - \gamma_+)}$ and is equal to $R(z_m) = \frac{-\gamma_-^2 + 4\Gamma(\gamma_+ - \Gamma)}{4(\Gamma - \gamma_+)}$. We have
$$
z_m \in [-1,+1] \iff \gamma_-^2 \leq 4(\Gamma - \gamma_+)^2.
$$
Now,
\begin{itemize}
\item[$-$]~ If $\Gamma\geq\gamma_+$, the extremum of $R$ is a minimum so $\max_{z\in[-1,+1]}R(z) = \max\{R(-1), R(1)\}$ and condition (\ref{radiusCondition2}) is $\max\{R(-1), R(1)\}\leq0$.
\item[$-$]~If $\Gamma\leq\gamma_+$, the extremum of $R$ is a maximum then
\begin{enumerate}
\item if $z_m \in [-1,+1]$ then condition (\ref{radiusCondition2}) is $\gamma_-^2 \leq 4\Gamma(\gamma_+ - \Gamma)$ (and $\Gamma\geq0$).
\item if $z_m \notin [-1,+1]$ then condition (\ref{radiusCondition2}) is $\max\{R(-1), R(1)\}\leq0$.
\end{enumerate}
\end{itemize}
Since $R(1) - R(-1) = 2\gamma_-$, the condition $\max\{R(-1), R(1)\}\leq0$ reads $|\gamma_-|\leq\gamma_+$.

So far we have proved the following.
\begin{itemize}
\item[$-$]~If $\Gamma \geq \gamma_+$, the dynamical map is P-divisible if and only if $|\gamma_-|\leq\gamma_+$.
\item[$-$]~If $\Gamma \leq \gamma_+$, the dynamical map is P-divisible if and only if $$
    \left\{
    \begin{array}{cc}
        |\gamma_-|\leq\gamma_+ & {\rm for}\ \gamma_-^2 \geq 4(\gamma_+ - \Gamma)^2  \\
        \gamma_-^2 \leq 4\Gamma(\gamma_+ - \Gamma) & {\rm for}\ \gamma_-^2 \leq 4(\gamma_+ - \Gamma)^2
    \end{array}
    \right.
    $$
\end{itemize}
Using $\gamma_+^2 - 4\Gamma(\gamma_+ - \Gamma) = (\gamma_+ - 2\Gamma)^2 \geq0$, we conclude that $\gamma_+^2 \geq  4\Gamma(\gamma_+ - \Gamma)$.
Therefore, in any case, the condition $|\gamma_-|\leq\gamma_+$ is satisfied when P-divisibility holds.
We thus have shown that the dynamical map is P-divisible if and only if
\begin{itemize}
    \item[$-$]~ $|\gamma_-|\leq\gamma_+$.
   \item[$-$]~If $\Gamma \leq \gamma_+$ and $\gamma_-^2 \leq 4(\gamma_+ - \Gamma)^2$, then $\gamma_-^2 \leq 4\Gamma(\gamma_+ - \Gamma)$.
\end{itemize}
Now notice that if $\Gamma \leq \gamma_+ \leq 2\Gamma$, then $4(\gamma_+ - \Gamma)^2 \leq 4\Gamma(\gamma_+ - \Gamma)$.
So in this case, P-divisibility holds if and only if $|\gamma_-|\leq\gamma_+$.
If $\gamma_+\geq 2\Gamma$, then $4(\gamma_+ - \Gamma)^2 \geq 4\Gamma(\gamma_+ - \Gamma)$ and, in this case, P-divisibility holds if and only if $\gamma_-^2 \leq 4\Gamma(\gamma_+ - \Gamma)$.
This concludes the proof.
\hfill\qedsymbol
\end{proof}

\section{Proof of Theorem~\ref{thm:breuer}}\label{appbreuer}

We set $|n\rangle =
\begin{pmatrix}
a\\
b
\end{pmatrix}$
and
$|m\rangle =
\begin{pmatrix}
c\\
d
\end{pmatrix}$
with $|a|^2 + |b|^2 =  |c|^2 + |d|^2=1$, $a\bar c + b\bar d = 0$.
Condition (\ref{Breuer1}) reads
\begin{eqnarray}
\label{Breuer2}
& & (\gamma_+ + \gamma_-)|a|^2|d|^2 + (\gamma_+ - \gamma_-)|b|^2|c|^2 \nonumber \\
& & +4(\Gamma - \frac{\gamma_+}{2})|b|^2|d|^2\geq0.
\end{eqnarray}
This condition only applies to moduli of complex numbers, so we can assume without losing generality that $a,b,c,d\in\R$\\
\noindent
Using $a^2 = 1 - b^2$ and $c^2 = 1 - d^2$, condition (\ref{Breuer2}) can be written only in terms of $b$ and $d$. We have
\begin{eqnarray*}
0 = (ac + bd)^2 &=& a^2c^2 + b^2d^2 + 2acbd\\
&=& (1 - b^2)(1 - d^2) + b^2d^2 - 2b^2d^2\\
&=& 1 - b^2 - d^2,
\end{eqnarray*}
from which we conclude that $b^2 + d^2 = 1$.
Condition (\ref{Breuer2}) can therefore be written as, $\forall d^2\in [0\,,1]$,
\begin{equation}
\label{Breuer3}
-4(\Gamma - \gamma_+)d^4 + 2(2\Gamma - 2\gamma_+ + \gamma_-)d^2 + \gamma_+ - \gamma_- \geq0.
\end{equation}\ \\
Suppose $\Gamma = \gamma_+$.
Condition (\ref{Breuer3}) then reads
\begin{equation}
\label{Breuer4}
2\gamma_-d^2 + \gamma_+ - \gamma_- \geq0,\ \forall d^2\in [0\,,1].
\end{equation}\ \\
\noindent
Condition (\ref{Breuer4}) implies that the affine function $P(X) = 2\gamma_- X + \gamma_+ - \gamma_-$ is positive over $[0\,,1]$.
This is equivalent to $P(0)\geq0$ and $P(1)\geq0$, that is, $\gamma_+ + \gamma_-\geq0$ and $\gamma_+ - \gamma_-\geq0$, or $|\gamma_-|\leq \gamma_+$.\\
Suppose now $\Gamma \neq \gamma_+$.
Condition (\ref{Breuer3}) has the form $P(X)\geq 0$ where $P$ is a polynomial of degree two in $X = d^2 \in [0\,,1]$. The coordinate of the extremum is $\alpha = \frac{1}{2} + \frac{\gamma_-}{4(\Gamma - \gamma_+)}$ and $\beta = \Gamma + \frac{\gamma_-^2}{4(\Gamma - \gamma_+)}$.
We get the following cases.
\begin{itemize}
\item[$-$]~ If $\Gamma > \gamma_+$, then the P-condition holds if and only if $P(0)\geq 0$ and $P(1)\geq 0$.
\item[$-$]~If $\Gamma < \gamma_+$ and $\alpha\in[0\,,1]$, then the P-condition holds if and only if $\beta\geq 0$.
\item[$-$]~ If $\Gamma < \gamma_+$ and $\alpha\notin[0\,,1]$, then the P-condition holds if and only if $P(0)\geq 0$ and $P(1)\geq 0$.
\end{itemize}
We have
\begin{eqnarray*}
P(0)\geq 0\ &\textrm{and}&\ P(1)\geq 0\nn\\ &\iff&\gamma_+ + \gamma_- \geq 0\ \textrm{and}\ \gamma_+ -\gamma_- \geq 0\\
&\iff& \gamma_+\geq0\ \textrm{and}\ \gamma_+ \geq |\gamma_-|\\
&\iff& \gamma_+ \geq |\gamma_-|.
\end{eqnarray*}
and
$$
\Gamma < \gamma_+\ \textrm{and}\ \beta\geq0 \iff \gamma_-^2 \leq 4\Gamma(\gamma_+ - \Gamma).
$$
Finally, we get
\begin{eqnarray*}
\Gamma < \gamma_+\ \textrm{and}\ \alpha\in[0\,,1] &\iff& \Gamma < \gamma_+\ \textrm{and}\ |\gamma_-| \leq 2(\gamma_+ - \Gamma)\\
&\iff& \Gamma < \gamma_+\ \textrm{and}\ \gamma_-^2 \leq 4(\gamma_+ - \Gamma)^2.
\end{eqnarray*}
Note that, in the second case, we necessarily have $P(0)\geq0$ and $P(1)\geq0$ since they are both smaller than or equal to $\beta\geq0$. Hence, the condition $\gamma_+ \geq |\gamma_-|$ is necessary in any case.
Our situation can also be reformulated as:\\
If $\Gamma < \gamma_+$ and $\alpha\in[0\,,1]$, then the P-condition holds if and only if $\beta\geq 0$ and $P(0)\geq0$ and $P(1)\geq0$.\\
\noindent
So far we have (including the $\Gamma = \gamma_+$-Case)

\begin{itemize}
\item[$-$]~ If $\Gamma \leq \gamma_+$ and $\gamma_-^2 \leq 4(\gamma_+ - \Gamma)^2$, then the P-condition holds if and only if $\gamma_-^2 \leq 4\Gamma(\gamma_+ - \Gamma)$ and $\gamma_+ \geq |\gamma_-|$.
\item[$-$]~ P-condition holds if and only if $\gamma_+ \geq |\gamma_-|$ otherwise.
\end{itemize}
Such conditions are the same as in the proof of Theorem~\ref{thm:tracenorm} (where it is also shown that they are equivalent to the conditions of Theorem~\ref{thm:tracenorm}).\hfill\qedsymbol

\section{Proof of constant ratio $\gamma_-(t)/\gamma_+(t)$ for the example in Sec.~\ref{sec5}. }\label{app:ratio}
In this section we derive the relation $\frac{\gamma_-(t)}{\gamma_+(t)} = -\tanh\left(\frac{ \omega_B\beta_B}{2}\right)$ valid for any initial thermal state of the bosonic mode $B$. First, we re-write the ratio as
\be
\frac{\gamma_-(t)}{\gamma_+(t)} = \frac{\gamma_1(t)/\gamma_2(t) - 1}{\gamma_1(t)/\gamma_2(t) + 1},
\ee
where $\gamma_1(t)$ and $\gamma_2(t)$ are given in Sec.~\ref{sec5}. We obtain
\bea
\frac{\gamma_1(t)}{\gamma_2(t)}=   \frac{\alpha(t)\dot\beta(t) -\beta(t)\dot\alpha(t) - \dot\beta(t)}{-\alpha(t)\dot\beta(t) +\beta(t)\dot\alpha(t) - \dot\alpha(t)}.
\eea
Then, from Eqs.~\eqref{eq:alpha} and \eqref{eq:beta} we can show that
\bea\label{eq:da}
&\dot \alpha(t)& = \sum_{n=1}^\infty p_n\frac{\Delta^2-\Omega_n^2}{\Omega_n} \cos \frac{\Omega_n t}{2} \sin \frac{\Omega_n t}{2}\\\label{eq:db}
&\dot \beta(t)& = \sum_{n=1}^\infty p_{n-1}\frac{\Delta^2-\Omega_n^2}{\Omega_n} \cos \frac{\Omega_n t}{2} \sin \frac{\Omega_n t}{2},
\eea
from which follow the identities:
\bea
&\dot \alpha(t)& = e^{-\omega_B\beta_B}\dot\beta(t)\nn\\
&\alpha(t)&\dot\beta(t) - \dot\alpha(t)\beta(t) =  \left(1-e^{-\omega_B\beta_B}\right)\dot\beta(t),
\eea
with $\beta_B = 1/k_BT_B$. This leads to
\be\label{eq:identities}
 \frac{\gamma_1(t)}{\gamma_2(t)} = e^{- \omega_B\beta_B},
\ee
giving finally
\be
\frac{\gamma_-(t)}{\gamma_+(t)} = -\tanh\left(\frac{ \omega_B\beta_B}{2}\right).
\ee

\end{document}